\documentclass[twocolumn,prd,showpacs,10pt]{revtex4}
\usepackage{graphicx}
\usepackage{epsfig}
\usepackage{amsmath, amssymb}

\def\OO{{\cal O}}
\def\dd{\partial}
\def\be{\begin{equation}}
\def\ee{\end{equation}}
\def\bea{\begin{eqnarray}}
\def\eea{\end{eqnarray}}
\def\ba{\begin{array}}
\def\ea{\end{array}}
\newcommand{\lsim}{\,\raise 0.4ex\hbox{$<$}\kern -0.8em\lower 0.62ex\hbox{$\sim$}\,}
\newcommand{\gsim}{\,\raise 0.4ex\hbox{$>$}\kern -0.7em\lower 0.62ex\hbox{$\sim$}\,}

\def\de{\mathrm{DE}}
\newcommand{\ca}{{c_a^2}}
\newcommand{\cs}{{\hat{c}_s^2}}
\newcommand{\ax}{a_\times}
\newcommand{\brho}{\bar{\rho}}
\newcommand{\dep}{\delta p}
\newcommand{\der}{\delta\!\rho}
\newcommand{\hdep}{\delta \hat{p}}
\newcommand{\hder}{\delta\!\hat{\rho}}
\newcommand{\bap}{\bar p}
\newcommand{\HH}{{\mathcal H}}

\newcommand{\eff}{{\mathrm{eff}}}
\newcommand{\nad}{{\mathrm{nad}}}
\newcommand{\rel}{{\mathrm{rel}}}
\newcommand{\cseff}{{\hat{c}_{s,\mathrm{eff}}^2}}
\newcommand{\csa}{{\hat{c}_{s,1}^2}}
\newcommand{\csb}{{\hat{c}_{s,2}^2}}
\newcommand{\p}{\phi}

\begin{document}

\title{Crossing the Phantom Divide}
\date{January 27, 2007}

\author{Martin Kunz}
\email{Martin.Kunz@physics.unige.ch}
\affiliation{D\'epartement de
Physique Th\'eorique, Universit\'e de
Gen\`eve, 24 quai Ernest Ansermet, CH--1211 Gen\`eve 4, Switzerland}
\author{Domenico Sapone}
\email{Domenico.Sapone@physics.unige.ch}
\affiliation{D\'epartement de
Physique Th\'eorique, Universit\'e de
Gen\`eve, 24 quai Ernest Ansermet, CH--1211 Gen\`eve 4, Switzerland}

\begin{abstract}
We consider fluid perturbations close to the ``phantom divide'' characterised
by $p=-\rho$ and discuss the conditions under which divergencies in the
perturbations can be avoided. We find that the behaviour of the perturbations
depends crucially on the prescription for the pressure perturbation $\dep$.
The pressure perturbation is usually defined using the dark energy rest-frame,
but we show that this frame becomes unphysical at the divide. If the pressure
perturbation is kept finite in any other frame, then the phantom divide can
be crossed. Our findings are important for generalised fluid dark energy used
in data analysis (since current cosmological data
sets indicate that the dark energy is characterised by $p\approx-\rho$
so that $p<-\rho$ cannot be excluded)
 as well as for any models crossing the
phantom divide, like some modified gravity, coupled dark energy and braneworld models.
We also illustrate the results by an explicit calculation for the ``Quintom'' case
with two scalar fields.
\end{abstract}

\maketitle

\section{Introduction}

The discovery by the supernova surveys \cite{sn1,sn2} 
that the expansion of the universe is currently accelerating came as a great
surprise to cosmologists. Within the standard cosmological
framework of a nearly isotropic and homogeneous universe
and an evolution described by General Relativity, this behaviour
requires a component with a negative pressure $p<-\rho/3$,
commonly dubbed {\em dark energy}. However, it is very difficult
to understand why the dark energy should appear at such a low
energy scale, and why it should start to dominate the overall
energy density just now. Explaining its nature is
correspondingly regarded as one of the most
important problems in observational cosmology.

Current limits on the equation of state parameter $w=p/\rho$
of the dark energy
seem to indicate that $p\approx -\rho$ \cite{cora,wlim}, 
sometimes even that $p<-\rho$ \cite{phantom}, often called
{\em phantom energy} \cite{caldwell}. Although there is
no problem to consider $w<-1$ for the background evolution, there
are apparent divergencies appearing in the perturbations 
when a model tries to cross the ``phantom divide'' $w=-1$\cite{crossing}. 
Even though this region may be unphysical at the
quantum level \cite{cht,Cline:2003gs}, it is still important to be able to probe
it, not least to test for alternative theories of gravity or
higher dimensional models which can give rise to an effective
phantom energy \cite{para,nojiri,vli,vik2,kosh}. It would certainly be unwise to build a strong
bias like $w\geq-1$ into our analysis tools as long as experiments
do not rule it out. In this paper we consider the evolution of
the perturbations for models where $w$ crosses $-1$. We find that
in many realistic cases the divergencies are only apparent and
can be avoided. At the level of cosmological first order perturbation 
theory, there is no fundamental limitation that prevents an effective
fluid from crossing the phantom divide.

The paper is organised as follows: We start with a short recapitulation
of the first order perturbations in fluids, which also serves to define
our notation. In section III we study the behaviour of barotropic fluids
close to $p=-\rho$, concluding that this class of fluids cannot cross the
phantom divide self-consistently. We allow for non-adiabatic perturbations
in section IV and show that the phantom divide can now be crossed as long
as we define the pressure perturbation in a frame that stays physical.
We then illustrate the results with the Quintom model of two scalar fields
before presenting our conclusions. The appendices finally discuss the
calculation of the effective Quintom perturbations in more detail.


\section{First order perturbations}

In this paper we use overdots to refer to derivatives with respect to
conformal time $\eta$ which is related to the physical time $t$ by
$dt=a d\eta$.
We will denote the physical Hubble parameter with $H$ and with $\HH$
the conformal Hubble parameter. For simplicity, we consider a flat 
universe containing only (cold dark) matter 
and a dark energy fluid, so that the Hubble parameter is given by
\be
H^2 = \left( \frac{1}{a} \frac{da}{dt} \right) ^2 = H_{0}^{2}\left[
  \Omega_m a^{-3}+\left( 1- \Omega_m \right) f(a) \right]
\ee
where $f(a)=\exp\left[ -3\int_1^a\frac{1+w(u)}{u}du \right]$,
implying that the scale factor today is $a_0=1$. We will assume
that the universe is filled with perfect fluids only,
so that the energy momentum tensor of each component is given by
\be
T^{\mu\nu}=\left( \rho+p\right) u^{\mu}u^{\nu} +p~g^{\mu\nu} .
\label{EMT}
\ee
where $\rho$ and $p$ are the density and the pressure of the 
fluid respectively and $u^{\mu}$ is the four-velocity.
This is potentially a strong assumption for the dark energy,
but a more general model should lead to more freedom in the
dark energy evolution, not less. 

We will consider linear perturbations about a spatially-flat background
model, defined by the line of element:
\bea
ds^{2} &=& a^{2} \left[ -\left( 1+2A\right) d\eta^{2}+2B_{i}d{\eta}dx^{i}+
  \right. \nonumber\\
&& \left. + \left( \left( 1+2H_{L}\right) \delta_{ij}+2H_{Tij} \right) dx_{i}dx^{j} \right]
\label{pert_0_ds}
\eea
where $A$ is the scalar potential; $B_{i}$ a vector shift; $H_{L}$ is the
scalar perturbation to the spatial curvature; $H_{T}^{ij}$ is the trace-free
distortion to the spatial metric, see e.g. \cite{hu_lecture} for more details.

The components of the perturbed energy moment tensor can be written as:
\bea
T_{0}^{0} &=& - \left( \bar\rho + \delta\rho \right) \\
T_{j}^{0} &=& \left( \bar\rho + \bar{p} \right) \left( v_{j} - B_{j} \right) \\
T_{0}^{i} &=& \left( \bar\rho + \bar p \right) v^{i} \\
T_{j}^{i} &=& \left( \bar{p} + \delta{p} \right) \delta_{j}^{i} + \bar{p}~\Pi_{j}^{i} .
\eea
Here $\bar\rho$ and $\bar p$ are the energy density and pressure of the
homogeneous and isotropic background universe,
$\der$ is the density perturbation, $\dep$ is the pressure perturbation,
$v^{i}$ is the vector velocity and $\Pi_{j}^{i}$ is the anisotropic stress
perturbation tensor \cite{hu_lecture}.

We want to investigate only the scalar modes of the perturbation equations.
We choose the Newtonian gauge (also known as the longitudinal gauge) which is
very simple for scalar perturbations because they are characterised by two
scalar potentials $\psi$ and $\phi$; the metric Eq. (\ref{pert_0_ds}) becomes:
\be
ds^{2} = a^{2} \left[ -\left( 1+2\psi \right) d\eta^{2} 
+ \left( 1-2\phi\right) dx_{i}dx^{i} \right]
\label{pert_newton_ds}
\ee
where we have set the shift vector $B_{i}=0$ and $H_{T}^{ij}=0$. The advantage of using the
Newtonian gauge is that the metric tensor $g_{\mu\nu}$ is diagonal and
this simplifies the calculations \cite{mabe}.

The energy-momentum tensor components in the Newtonian gauge become:
\bea
T_{0}^{0} &=& -\left( \brho + \delta\rho \right) \label{00_EM}\\
ik_i T_{0}^{i} &=& -ik_i T_{i}^{0} = \left(\brho + \bap \right) \theta  \label{0i_EM}\\
T_{j}^{i} &=& \left( \bar p + \delta p \right) \delta_{j}^{i}  \label{ij_EM}
\eea
where we have defined the variable $\theta=ik_j v^j$ which represents the divergence
of the velocity field and we have also assumed that the anisotropic stress vanishes,
$\Pi_{j}^{i}=0$ (implying $\phi=\psi$).

The perturbation equations are \cite{mabe}:
\bea
\dot\delta &=& -\left( 1+w \right) \left( \theta - 3\dot\psi \right)
-3\frac{\dot a}{a} \left( \frac{\delta p}{\bar\rho} - w\delta \right) \label{d_pert}\\
\dot\theta &=& -\frac{\dot a}{a} \left( 1-3w \right) \theta -
\frac{\dot{w}}{1+w}\theta  + \nonumber \\
&& +k^{2}\frac{\delta{p}/\bar\rho}{1+w} + k^{2}\psi \label{t_pert} .
\eea
As $w \rightarrow -1$ the terms containing $1/(1+w)$ will generally
diverge. This can be avoided by replacing $\theta$ with a new variable
$V$ defined via $V=\left( 1+w \right) \theta$. This corresponds to rewriting
the $0i$ component of the energy momentum tensor as 
$ik_j T_{0}^{j}=\brho V$ which avoids problems if $T_{0}^{j}\neq0$ when
$\bap=-\brho$. Replacing the time derivatives by a derivative with respect
to the scale factor $a$ (denoted by a prime), we obtain:
\bea
\delta' &=& 3(1+w) \psi' - \frac{V}{Ha^2} 
- 3 \frac{1}{a}\left(\frac{\dep}{\bar\rho}-w \delta \right) \label{eq:delta} \\
V' &=& -(1-3w) \frac{V}{a}+ \frac{k^2}{H a^2} \frac{\dep}{\bar\rho}
+(1+w) \frac{k^2}{Ha^2} \psi .  \label{eq:v}
\eea
In this form everything looks perfectly finite even at $w=-1$.
But in order to close the system, we need to give an expression
for the pressure perturbations. A priori this is a free choice
which will describe some physical properties of the fluid. In
the following we will consider several possible choices and
discuss how they influence the behaviour of the fluid at the
phantom divide. We will see how the specification of
$\delta p$ determines if a fluid can cross the divide or not.

In addition to the fluid perturbation equations we need to add
the equation for the gravitational potential $\psi$. If there are several fluids
present, then the evolution of each of them will be governed by their own set
of equations for their matter variables $\{\delta_i,V_i\}$, linked by a
common $\psi$ (which receives contributions from all the fluids)\footnote{We assume 
that there are no couplings beyond gravity between the fluids.}. 
We use \cite{mabe},
\be
\psi' = -\frac{1}{2a} \left(\frac{\sum_i \delta_i \brho_i}{\sum_i \brho_i}\right)
-\left(\frac{k^2}{3 H^2 a^2}+1\right) \frac{\psi}{a} .
\label{eq:psi}
\ee

The evolution of a fluid is therefore governed by Eqs.~(\ref{eq:delta}) and 
(\ref{eq:v}), supplemented by a prescription for the internal physics (given
by $\dep$) and the external physics (through $\psi$ and $\psi'$). There are
two points worth emphasising. Firstly, the dark energy fluid perturbations
cannot be self-consistently set to zero if $w\neq-1$. Even if $\delta=V=0$
on some initial hypersurface, it is unavoidable that perturbations will
be generated by the presence of $\psi$. As this function describes physics
external to the fluid, it cannot be controlled directly. Conceivably $\dep$
could be chosen to cancel the external source of either $\delta$ or $V$,
but not of both. Even then such a $\dep$ would have to be incredibly 
fine-tuned as $\psi$ and $\psi'$ depend on the evolution of the other
fluids as well.

Secondly, we see that as $w\rightarrow-1$ the external sources are turned
off. A fluid can therefore mimic a cosmological constant as then (and only
then) $\delta=V=\dep=0$ is a solution. This corresponds to an energy momentum
tensor (\ref{EMT}) with $T_{\mu\nu} = \bap g_{\mu\nu}$, so that 
$\Lambda \sim \brho$. However, in general the perturbations do not vanish even
if $w=-1$. A perfect fluid is therefore in general {\em not} a cosmological
constant even if $\brho=-\bap$.

To calculate perturbations in different gauges we need to introduce 
the coordinate transformation:
\bea
\eta &=& \tilde{\eta} + T  \\
x^i &=& \tilde{x}^i + L^i .
\eea
the gauge transformation of the matter variables is then:
\bea
\tilde{\delta\rho} &=& \delta\rho +3\frac{\dot a}{a}\left( 1+w \right) \bar\rho T
\label{d_rho_gauge} \\
\frac{\tilde{\delta p}}{\bar p} &=& \frac{\delta p}{\bar p} +3\frac{\dot a}{a}
\left( 1+w \right) \frac{c_{a}^{2}}{w} T \label{d_p_gauge}\\
\tilde{u} &=& u + \dot L \label{d_v_gauge} 
\eea
where we have introduced a new quantity
$c_a^2=\dot{\bar{p}}/\dot{\bar{\rho}}$, called adiabatic sound speed.


\section{Barotropic fluids}

\begin{figure}
\epsfig{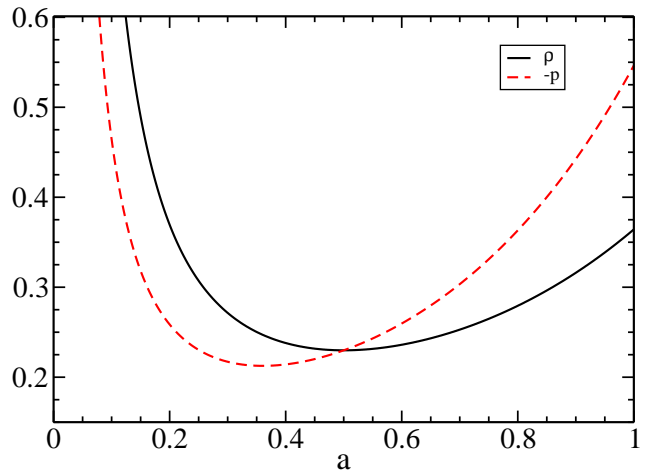}
\caption{Energy density (black solid line) and pressure (red dashed line) as
  functions of the scale factor. 
  $w$ crosses $-1$ at $\ax=0.5$, so that $\rho$ is minimal at this point
  while $p$ decreases monotonically there (but has a maximum earlier). 
  }
\label{fig-1a}
\end{figure}

\begin{figure}
\epsfig{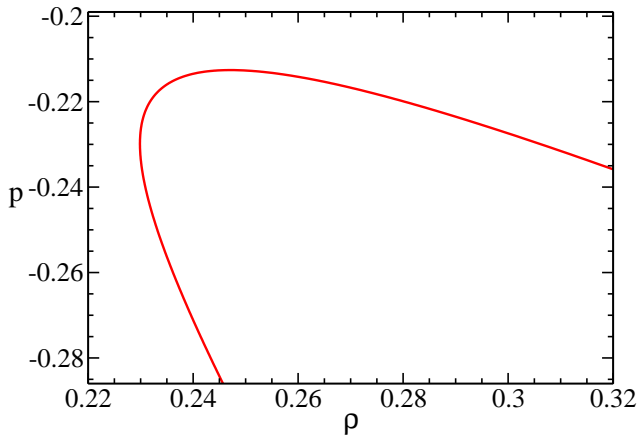}
\caption{The pressure as function of the energy
  density. The graph
  of $p(\rho)$ shows that $p$ is not a single valued function of the energy
  density, and that there is a point with an infinite slope (corresponding to
  the minimum of $\rho$ and the divergence of $\ca$). The point where $p(\rho)$
  has a zero slope corresponds to the maximum of $p$ and a vanishing $\ca$.
  }
\label{fig-1b}
\end{figure}

We define a fluid to be barotropic if the pressure $p$ depends strictly
only on the energy density $\rho$: $p=p(\rho)$. These fluids 
have only adiabatic perturbations, so that they are often called adiabatic.
We can write their pressure as
\be
p(\rho) = p(\bar{\rho}+\delta\rho) 
= p(\bar{\rho}) + \left.\frac{dp}{d\rho}\right|_{\bar{\rho}} \delta\rho
+ O\left((\delta\rho)^2\right).
\label{eq:baro_exp}
\ee
Here $p(\brho) = \bap$ is the pressure of the isotropic and homogeneous
part of the fluid. Introducing
$N\equiv \ln a$ as a new time variable, we can rewrite the background 
energy conservation equation, $\dot{\brho} = -3 \HH (\brho+\bap)$ in terms of $w$,
\be
\frac{dw}{dN} = \frac{d\brho}{dN} \frac{dw}{d\brho}
= -3 (1+w) \brho  \frac{dw}{d\brho} .
\label{eq:w_baro}
\ee
We see that the rate of change of $w$ slows down as $w\rightarrow -1$, and
$w=-1$ is not reached in finite time \cite{vik1}, except if $dw/d\brho$ diverges
(or $\brho$, but that would lead to a singular cosmology).
The physical reason is that we demand
$p$ to be a unique function of $\rho$, but at $w=-1$ we find that $\dot{\brho}=0$.
If the fluid crosses $w=-1$ the energy density $\rho$ will first decrease
and then increase again, while the pressure $p$ will monotonically
decrease (at least near the crossing), Fig.~\ref{fig-1a}. It is therefore impossible 
to maintain a one-to-one relationship between $p$ and $\rho$, 
see Fig.~\ref{fig-1b} (notice that the maximum of $p$ and the minimum
of $\rho$ do not coincide).

An example that can potentially cross $w=-1$ is given by
\be
\left(1+w(\rho)\right)^2 = C^2 (\brho-\brho_\times) .
\label{eq:ex_baro}
\ee
Starting with $w>-1$ and $\brho>\brho_\times$ both will decrease until
$w=-1$ and $\brho=\brho_\times$. If it is possible to switch from the
branch with $w>-1$ to the other one at this point, then
$\brho$ can start to grow again while $w$ continues to decrease.
Using the evolution equation (\ref{eq:w_baro}) for $w$, we
find
\be
\frac{dw}{dN} = -\frac{3}{2} \left[ C^2 \brho_\times + (1+w)^2 \right]
\ee
so that $dw/dN \simeq -3/2 C^2 \brho_\times$ near the crossing. The full
solution (with $N=N_\times$ at $w=-1$) is
\be
w(N) = -1 - C \sqrt{\brho_\times} \tan\left(\frac{3 C \sqrt{\brho_\times}}{2} 
(N-N_\times) \right) ,
\ee
clearly not a realistic solution for our universe as there are
divergencies in the finite past and future. Of course we could modify
the example (\ref{eq:ex_baro}), but we are mostly interested at the
behaviour near the crossing. In our case $1+w(a) \propto -(a-\ax)$, ie.
we observe a linear behaviour. This is actually the limiting case:
if we choose $(1+w) \propto (\brho - \brho_\times)^\nu$ close to the
crossing, we find
\be
\frac{d(1+w)}{dN} \simeq - C (1+w)^{(2\nu-1)/\nu}
\ee
for small $|1+w|$. Correspondingly we cross with zero slope for $\nu>1/2$
and with infinite slope for $\nu<1/2$. However, if $dw/dN=0$ at $w=-1$ then
the system will turn into a cosmological constant and stay there forever.
On the other hand, this is unstable against small perturbations towards
$\bap<-\brho$, so that the crossing might eventually be completed anyway.

Let us have a closer look at the perturbations. The second term
in the expansion (\ref{eq:baro_exp}) can be re-written as
\be
\left.\frac{dp}{d\rho}\right|_{\bar{\rho}}
= \frac{\dot{\bar{p}}}{\dot{\bar{\rho}}} = w - \frac{\dot{w}}{3 \HH (1+w)} 
\equiv c_a^2
\ee
where we used the equation of state and the conservation
equation for the dark energy density in the background.
We notice that the adiabatic sound speed $c_a^2$ will necessarily diverge
for any fluid where $w$ crosses $-1$. 

For a perfect barotropic fluid the adiabatic sound speed $\ca$ turns out to
be the physical propagation speed of perturbations. It should therefore never 
be larger than the speed of light, otherwise our theory becomes acausal
\cite{ah,bcd}. The condition
$\ca\le1$ for $(1+w)>0$ (our point of departure) is equivalent to
\be
-\frac{dw}{dN} \le 3 (1+w) (1-w) .
\ee 
No barotropic fluid can therefore pass through $w=\pm 1$ from above without
violating causality. Even worse, the pressure perturbation
\be
\dep = c_{a}^{2} \der = \left( w - \frac{\dot{w}}{3 \HH (1+w)} \right) \der
\ee
will necessarily diverge if $w$ crosses $-1$ and $\der\neq0$.
Using the gauge transformation Eqs.~(\ref{d_rho_gauge}) and (\ref{d_p_gauge}) we
see that the relation between pressure and energy density perturbations
is gauge invariant. If the pressure perturbation diverges in one frame, it
will diverge in all frames. The only possible way out would be to force
$\der\rightarrow0$ fast enough at the crossing. Let us study the behaviour
close to $w=-1$ in some detail now:

First we look only at the dominant contribution to the right hand side of
Eq.~(\ref{eq:delta}). Near $w=-1$ this is clearly:
\be
\delta' = -\frac{3}{a} \ca \delta = + \frac{w'}{1+w} \delta .
\label{eq:dcross1}
\ee
Assuming that near the crossing the equation of state behaves
like $1+w \approx \lambda (a-\ax)^\alpha$ with $\alpha>0$ 
we find the solution:
\be
\delta \propto (a-\ax)^\alpha
\ee
(independent of $\lambda$ which drops out of the equation).
This solution goes to zero at the crossing. If the sign had
been different, we would have found instead the solution
$\delta\propto(a-\ax)^{-\alpha}$ which diverges.

\begin{figure}
\epsfig{figure=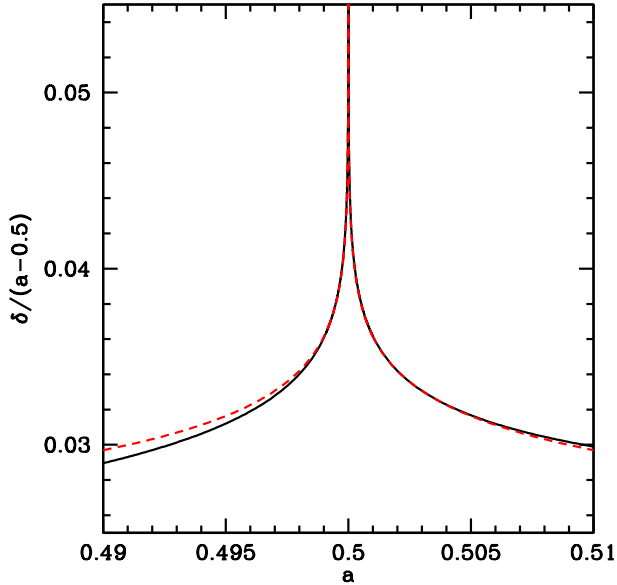,
  width=3.3in}
\caption{This figure shows the logarithmic divergence of
$\delta/(a-\ax)$ near $\ax=0.5$. The black curve is the
numerical solution while the red dashed line shows the approximation
given by Eq.~(\ref{eq:ad_log_div}) for $\alpha=1$. The approximate
formula describes the structure of the divergence quite well. The
pressure perturbation $\dep$ exhibits a very similar divergence in
this case.}
\label{fig_log_diverge}
\end{figure}

The pressure perturbation behaves like
\be
\dep \propto \brho \left( \frac{\alpha a}{3} (a-\ax)^{\alpha-1}
+ \OO((a-\ax)^{\alpha}) \right) . \label{eq:dep1}
\ee
We need $\alpha\geq1$, otherwise the pressure perturbation will
diverge as a power-law at the crossing. Unfortunately this condition
is not sufficient. Looking at the second perturbation equation
(\ref{eq:v}) we see that although nothing diverges, there is no
reason for $V$ to go to zero at the crossing, and in general
it will not, except if we fine-tune it with infinite precision.
As $\delta$ vanishes at crossing and can so potentially cancel the 
divergence in the sound speed, this term is no longer necessarily dominant
in the differential equation for $\delta$, Eq.~(\ref{eq:dcross1}).
We also need to take into account the other contributions.
The term
containing $\psi'$ is sufficiently suppressed by the $1+w$
factor to neglect it. Taking $V$ to be constant at $\ax$ (to lowest order) 
$V(\ax)=V_\times$, we find:
\be
\delta' = - C_\times  + \frac{w'}{1+w} \delta
\ee
with $C_\times = V_\times/(H(\ax)\ax^2)$. The solution to this
equation is:
\be
\delta = \left\{
\begin{array}{ll} (a-\ax) \left(\delta_\times - C_\times \log|a-\ax| \right)  & \alpha=1 \\
\delta_\times (a-\ax)^\alpha + (a-\ax) C_\times/(1-\alpha) & \alpha \neq 1 \end{array}
\right.
\label{eq:ad_log_div}
\ee
We notice that for $\alpha\geq1$ the density perturbations vanish
at crossing. However, for $\alpha=1$ they do not vanish fast enough.
Instead of the behaviour given in Eq.~(\ref{eq:dep1}), the pressure
perturbation exhibits now a logarithmic divergence since the sound
speed cancels the factor $(a-\ax)$. Even though our derivation here
is not rigorous, numerical calculations confirm this behaviour,
also if the background is not completely matter dominated, see
Fig.~\ref{fig_log_diverge}. Although a full solution may be possible
in some cases, it would turn out to be a special function, obscuring
the structure of the result while still showing essentially the same
simple behaviour.

We find therefore three possible behaviours for the perturbations,
depending on how $w$ crosses the phantom divide:
\begin{enumerate}
\item $\alpha<1$: In this case $w$ crosses $-1$ with an infinite slope,
leading to a power-law divergence of the pressure perturbation.
\item $\alpha=1$: $w$ crosses $-1$ with a finite, but non-zero slope.
$\der$ vanishes, $V$ stays finite and generally non-zero and $\dep$
diverges logarithmically.
\item $\alpha>1$: $w$ crosses $-1$ with a zero slope. $\der$ vanishes,
$V$ and $\dep$ stay finite and generally non-zero.
\end{enumerate}

The only acceptable case is the last one, but as discussed at the
beginning of this section, we expect the system to get stuck at $w=-1$
and to never cross. Barotropic perfect fluids therefore fail to cross
either at the background or perturbation level. Even if the fluctuations
can lead to a crossing with zero slope due to the instability of the
cosmological constant solution in some cases, 
the model is still not realistic. The
perturbations seem to propagate acausally, and generically after
the transition $\ca<0$ for a period, leading to classical instabilities
with exponential growth of the perturbations. In the next section
we will relax the barotropic assumption, which allows for entropy
perturbations that can stabilise the system and keep the sound speed
below the speed of light.

\section{Non-adiabatic fluids \label{sec:non_ad}}

The discussion of the barotropic fluid shows that we have to violate
the constraint that $p$ be a function of $\rho$ alone. At the level
of first order perturbation theory, this amounts
to changing the prescription for $\dep$ which now becomes an arbitrary
function of $k$ and $t$. This problem is conceptually similar to
choosing the background pressure $\bap(t)$, where the conventional
solution is to compare the pressure with the energy density by
setting $\bap(t) = w(t) \brho(t)$. In this way we avoid having to deal
with a dimensionfull quantity and can instead set $w$, which has no units
(up to a factor of the speed of light) and so is generically of order
unity and often has a simple form, like $w=1/3$ for a radiation fluid or
$w=-1$ for a cosmological constant.

It certainly makes sense to try a similar approach for the pressure
perturbation. However, there are two relevant variables that we could
compare $\dep$ to, the fluid velocity $V$ and the perturbation in the
energy density $\der$. Clearly it would be counterproductive to replace
a single free function by two free functions, and it would lead to
degeneracies between the two. Another problem is that the perturbation
variables depend on the gauge choice. But in this case the two problems
cancel each other, leading to a simpler solution: Going to the rest-frame
of the fluid both fixes the gauge and renders the fluid velocity
physically irrelevant, so that we can now write \cite{bedo}:
\be
\hdep = \cs \hder , \label{eq:csrest}
\ee
where a hat denotes quantities in the rest-frame. The physical
interpretation is that $\cs(k,t)$ is the speed with which fluctuations
in the fluid propagate, ie. the sound speed. Again, some physical
models lead to simple prescriptions for the sound speed. The barotropic
models discussed in the last section have $\cs(k,t) = \ca(t)$, and the
perturbations in a scalar field correspond (to linear order) exactly
to those of a fluid with $\cs(k,t)=1$.

The rest-frame is chosen
so that the energy-momentum tensor looks diagonal to an observer in
this frame. In terms of the gauge transformations
this amounts to choosing $B=\theta_{\de}$. 
However, we notice immediately a potential flaw in this prescription close
to $w=-1$: The off-diagonal entries of the energy-momentum tensor are
actually $(\brho+\bap)\theta$ so that demanding $\theta-B=0$ is a stronger
condition than required. In other words, the condition to be at rest with respect to
the flow of $\brho+\bap$ cannot be maintained at $\brho+\bap=0$. However,
as $\brho>0$, we could define instead the sound speed in a frame where
there is no flow of energy density.

To see this, let us calculate the pressure perturbation defined by
Eq.~(\ref{eq:csrest}) in the conformal Newtonian frame, following \cite{bedo}:
Breaking the single link between $\rho$ and $p$ amounts to the introduction
of entropy perturbations. A gauge invariant entropy perturbation variable 
is $\Gamma \equiv \frac{\delta{p}}{\bar{p}}-\frac{c_a^2}{w}
\frac{\der}{\brho}$ \cite{KS,bedo}. By using
\be
w \Gamma = \frac{\dot{p}}{\rho} \left(\frac{\dep}{\dot{p}}-
\frac{\der}{\dot{\rho}}\right)
\ee
and the expression for the gauge transformation of $\der$ \cite{KS},
\be
\hder = \der + 3 \HH \brho \frac{V}{k^2}
\ee
we find that the pressure perturbation is given by
\be
\dep = \cs \der + 3 \HH \left(\cs - \ca\right) \brho \frac{V}{k^2} .
\label{eq:dp_rest}
\ee
As $\ca \rightarrow \infty$ at the crossing, it is impossible that all
other variables stay finite except if $V\rightarrow 0$ fast enough.
Again we will show that this is not in general the case, except if
$\cs\rightarrow0$ or $w'\rightarrow0$ at crossing. However, we will then see that this
is not required, indeed we will argue in the next section that the 
more generic solution
is to let $\cs$ diverge at the crossing in order to cancel the
divergence of $\ca$!

To this end, we consider again the structure of the perturbation
equations near $w=-1$. Inserting the expression for $\dep$ into our
system of perturbation equations we find
\bea
\delta' &=& 3(1+w) \psi' - \left(\frac{k^2}{Ha^2}+3H(\cs-\ca) \right) \frac{V}{k^2} \nonumber  \\
    &&-\frac{3}{a}\left(\cs-w\right)\delta \label{eq:pert_cs_d}\\
\frac{V'}{k^2} &=& -\frac{1+3(\ca-w-\cs)}{a} \frac{V}{k^2} + \nonumber  \\
 &&+\frac{1+w}{H a^2} \psi + \frac{\cs}{H a^2} \delta . \label{eq:pert_cs_v}
\eea
The spoiler here is the continued presence of $\ca$ which we know to diverge at
crossing. Let us start by considering a finite and constant $\cs$. For all
reasonable choices of $k$ the dominant term 
in the $V$ equation will be the one containing $\ca$. We
find that this time the velocity perturbation is driven to zero at crossing:
\be
V' = \frac{w'}{1+w} V .  \label{eq:cs0cross}
\ee
Proceeding as in the last section, we find that now $V\propto (a-\ax)^\alpha$,
cancelling the divergence in $\ca$ for $\alpha\geq1$. Now
$\der$ will in general not vanish at crossing, and we have to include that term
in the differential equation for $V$. The term with $\psi$ is
suppressed by $1+w$ and is of higher order. As in the last section for
$\delta$, the solutions for $V$ are now to lowest order:
\be
V = \left\{
\begin{array}{ll} (a-\ax) \left(V_\times - D_\times \log|a-\ax| \right)  & \alpha=1 \\
V_\times (a-\ax)^\alpha + (a-\ax) D_\times/(1-\alpha) & \alpha \neq 1 \end{array}
\right. \label{eq:V_cross}
\ee
with $D_\times$ being $\cs k^2 \delta / (Ha^2)$, evaluated at crossing.

Again there are no divergences appearing in the energy momentum tensor {\em only}
if $\alpha>1$, ie. if $w$ crosses $-1$ with a zero slope. A possible way to get
around this fine-tuning is to demand that $\cs=0$ at crossing, as in this case
the logarithmically divergent term disappears. We also notice that the usual
velocity perturbation $\theta \equiv V/(1+w)$ does diverge in all cases,
either logarithmically if $\alpha=1$ or as $(a-\ax)^{1-\alpha}$ if $\alpha>1$.
For an observer in the rest-frame where $B=\theta$ this means that the
metric perturbations become large -- at crossing the metric is even singular.
At the very least, perturbation theory is no longer valid for such an
observer.

Another way to see that the rest-frame is ill-defined is to look at the
energy momentum tensor (\ref{EMT}). For $p=-\rho$ the first term disappears,
leaving us with $T^{\mu\nu}=p~g^{\mu\nu}$. Normally the four-velocity 
$u^\mu$ is the time-like eigenvector of the energy momentum tensor, but
now suddenly all vectors are eigenvectors. The problem of fixing a unique
rest-frame is therefore no longer well-posed.

However, {\em by construction} the pressure perturbation looks perfectly
fine for precisely {\em the observer in the rest-frame}, 
as $\hdep = \cs \hder$ does {\em not}
diverge. Our prescription for the pressure perturbations has singled out
the one frame which we cannot use for fluids crossing the phantom divide.
The reason is that the gauge transformation
relating the pressure perturbations in the different gauges is: 
\be
\hdep = \dep + 3\HH \brho \ca \frac{V}{k^2} .
\label{eq:dep_gauge}
\ee
If $V$ does not vanish fast enough at the crossing then the pressure
perturbation has to diverge in at least one frame. As we have just
discussed, the dark-energy rest-frame becomes unphysical at crossing. It is
clearly better to specify a finite pressure perturbation in a different
frame.

\begin{figure}
\epsfig{figure=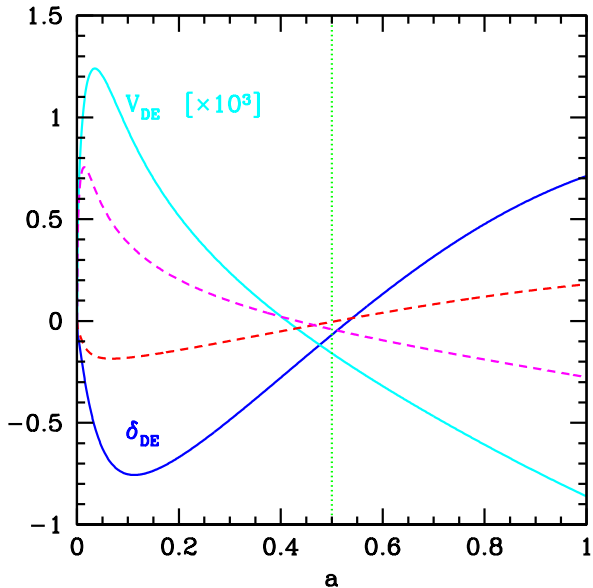,
  width=3.3in}
\caption{This figure shows the density contrast $\delta_\de$
(red and blue lines, negative for small scale factor $a$) and
the velocity perturbation $V_\de$ (cyan and magenta lines,
positive for small $a$) for a dark energy component with
a pressure perturbation given in the conformal Newtonian
gauge through $\gamma$ (as defined in Eq.~(\ref{d_p_gamma})).
The solid lines show the results for $\gamma=0.2$ and the
dashed lines for $\gamma=1$.The crossing of
the phantom divide at $\ax=0.5$ (dotted green vertical line)
is not apparent in these figures, everything is finite.}
\label{fig-2a}
\end{figure}

\begin{figure}
\epsfig{figure=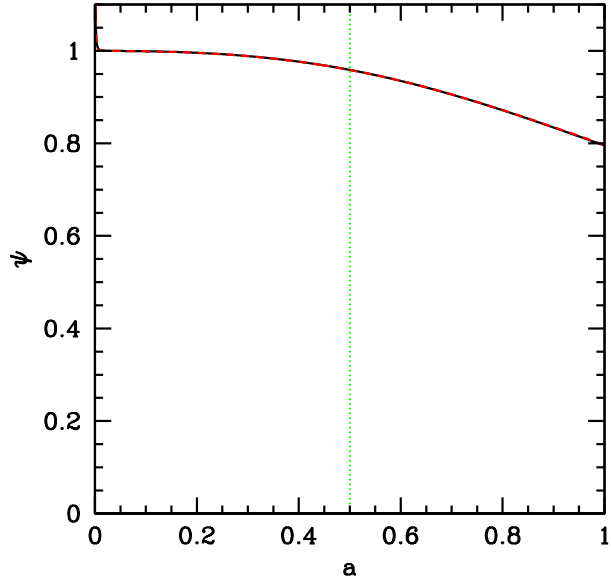,
  width=3.3in}
\caption{The gravitational potential $\psi$ in the same scenario 
as Fig.~\ref{fig-2a} (with $\gamma=0.2$ for the black solid line
and $\gamma=1$ for the red dashed line). Again, the crossing of
the phantom divide at $\ax=0.5$ (dotted green vertical line)
is not apparent. The gravitational potential is constant during
matter domination (small $a$) and starts to decay as the dark
energy begins to dominate and the expansion rate of the universe
accelerates.  The contribution of the dark energy perturbations
to $\psi$ is very small ($\lsim 0.1\%$) 
so that there is no visible difference for the two choices of $\gamma$.}
\label{fig-2b}
\end{figure}

One problem is to find a way of characterising the pressure perturbations
in a physical way -- in the
Quintom example of the next section, we find that the additional
contributions diverge but in such a way that we end up with a finite
result for $\dep$. As another example, we just choose $\dep$ proportional
to $\der$,
\be
\dep(k,t) = \gamma\, \der(k,t) \label{d_p_gamma}
\ee
in the conformal Newtonian gauge. This will work as long as $\der\neq 0$,
otherwise it forces $\dep$ to vanish in the same place as $\der$, which is
not general enough. If we insert this expression into
eqs. (\ref{eq:delta}) and (\ref{eq:v}), we obtain:
\bea
\delta' &=& 3(1+w) \psi' - \frac{V}{Ha^2} 
- 3 \frac{1}{a}\left( \gamma-w  \right) \delta \label{prime} \\
V' &=& -(1-3w) \frac{V}{a}+ \frac{k^2}{H a^2} \gamma \delta
+(1+w) \frac{k^2}{Ha^2} \psi \label{second}. 
\eea
As $w\rightarrow -1$ none of the terms diverge, so that $\delta$
and $V$ stay in general finite and non-zero at crossing. We show in
Figs.~\ref{fig-2a} and \ref{fig-2b} a numerical example for two choices of $\gamma$
where it is impossible to see that the phantom divide has been 
crossed at $\ax=0.5$.

It is of course possible to express $\gamma$ in terms of $\cs$,
and $\cs$ in terms of $\gamma$. Using the expression for the pressure
perturbation in the rest-frame given by Eq.~(\ref{eq:csrest}) and the gauge
transformation given by Eqs.~(\ref{d_rho_gauge}) and (\ref{d_p_gauge}), we
obtain
\be
\gamma \delta\rho+3\HH c_a^2\brho\frac{V}{k^2}=\cs\left( \der+3\HH \brho\frac{V}{k^2} \right) .
\ee 
In general $\gamma$ will therefore be scale dependent even if $\cs$ is not
(even though of course $\cs$ will also in general depend on $k$ and $t$),
or vice versa. Also, to reproduce the evolution with finite $\gamma$ shown
in Figs.~\ref{fig-2a} and \ref{fig-2b} we would have to substitute a divergent $\cs$ to cancel
the divergence in $\ca$ (cf. figure \ref{fig:quint_cs} which shows how
the apparent sound speed diverges in the Quintom example). 
We also notice that on very small scales where
$k\gg1$ we find $\gamma\approx\cs$ for finite $\ca$, which is the usual result
that on small scales gauge differences become irrelevant (for physical gauges).

Finally we would like to emphasise again that for a general, finite
pressure perturbation in any gauge except the fluid rest-frame, there
is no problem with the perturbation evolution across the phantom divide.
Also, in general the perturbations, including $V$, will not vanish at
crossing.

\section{The Quintom model as explicit example}

To clarify some of the above points we consider an explicit example
of a model crossing the phantom divide, the Quintom model 
\cite{quintom1,quintom2},
and compute the pressure perturbation {\em ab initio} at crossing
(see also \cite{qpert}).

The original Quintom model considered two scalar fields. For us it is
advantageous to use instead two fluids with a constant and equal rest-frame
sound speed $\cs=1$. At the level of first order perturbation theory the
two models are exactly equivalent, see appendix \ref{app:scalar}.
As in the original model, we use two fluids with constant equations of
state parameters $w_1$ and $w_2$. We will have to map the two fluids
onto a single effective fluid. To this end we define the effective parameters
in such a way that the effective energy momentum tensor is the sum of the
two fluid energy momentum tensors. This leads to:
\bea
\brho_\eff &=& \brho_{1} + \brho_{2}\\
\bap_\eff &=& \bap_{1} + \bap_{2}\\
w_\eff &=& \frac{w_{1}\brho_{1} + w_{2}\brho_{2}}{\brho_{1}+\brho_{2}}\\
\delta_\eff &=& \frac{\brho_{1}\delta_{1} + \brho_{2}\delta_{2}}{\brho_{1}+\brho_{2}} \\
\theta_\eff&=&\frac{\left( 1+w_{1} \right)\brho_{1}\theta_{1}+\left( 1+w_{2} \right)\brho_{2}\theta_{2}}{\left( 1+w_{1} \right)\brho_{1}+\left( 1+w_{2} \right)\brho_{2}} .
\eea

Re-expressing the perturbation equations in these variables we find precisely
the normal perturbation equations for a single fluid (\ref{eq:delta}) and 
(\ref{eq:v}) with $\bap_\eff$
and $\dep_\eff$ replacing $p$ and $\dep$. 
$\bap_\eff$ is simply given by $w_\eff \brho_\eff$, but
the density perturbation is more involved. Starting from
$\dep_\eff = \dep_1 + \dep_2$ we can write it as
\bea
\dep_\eff&=& \cseff \der_\eff+\dep_{\rel} +\dep_{\nad} \nonumber\\
&&+ 3\HH\left(\cseff-\ca \right) \brho_\eff \frac{ V_\eff}{k^2}
\label{eq:dp_quintom}
\eea
where the (effective) adiabatic sound speed is as always 
$\ca = \dot{\bap}_\eff/\dot{\brho}_\eff$ and
where $\cseff=1$ for two scalar fields.
The other terms are the relative pressure perturbation
\be
\dep_{\rel} = \frac{\left( w_{2} - w_{1} \right) \dot\brho_{1}
  \dot\brho_{2}}{3\HH \dot{\brho}_\eff} S_{12}
\ee
given by the relative density perturbation of the two scalar fields,
\be
S_{12}=\frac{\delta_{1}}{1+w_{1}}-\frac{\delta_{2}}{1+w_{2}} \label{entropy_pert}
\ee
(which corresponds to a gauge invariant relative entropy 
perturbation\cite{MaWa})
as well as the non-adiabatic term
\be
\dep_{\nad} = -\left(w_{2}-w_{1}\right)
\frac{\dot\brho_{1}\dot\brho_{2}}{3\HH\dot\brho_\eff}
\left[S_{12}+ \frac{3\HH}{k^2}\Delta\theta_{12}\right]
\ee
given by the relative motion of the two scalar fields 
with $\Delta\theta_{12}=\theta_1-\theta_2$
(see appendix \ref{app:2fluids}). 

We notice that the relative perturbations act as internal perturbations which
couple to the effective variables purely through the behaviour of the pressure
perturbation. Understanding the physics behind the dark energy in such a case
will therefore require a precise measurement of the pressure perturbations
as well as their careful analysis, to uncover the different internal
contributions.

\begin{figure}
\epsfig{figure=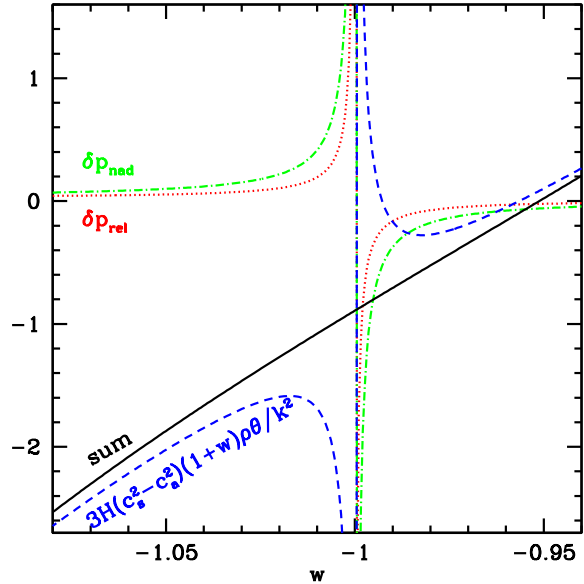,
  width=3.3in}
\caption{This figure shows the different divergent contributions to the pressure
perturbation, Eq.~(\ref{eq:dp_quintom}), multiplied by $10^9$. The relative
pressure perturbation is shown as red dotted line, the non-adiabatic pressure
perturbation as green dash-dotted line and the contribution from the gauge transformation
to the conformal Newtonian frame as blue dashed line. Each of the contributions diverges at
the phantom crossing, but their sum (shown as black solid line), and so $\dep$, stays finite.}
\label{fig:quintom1}
\end{figure}

Even though the effective sound speed $\cseff=1$ in Eq.~(\ref{eq:dp_quintom})
is finite in the rest-frame, the transformation
to any other frame will lead to a divergence as discussed in the
previous section. This divergence then needs to be cancelled by the
two additional terms, $\dep_{\rel}$ and $\dep_{\nad}$. In our
example, {\em both} diverge, cancelling together the divergent
contribution from the singular gauge transformation as well as
their own divergencies, see Fig.~\ref{fig:quintom1}.  This behaviour, which
looks extremely fine-tuned, is automatically enforced in this model.
Such a cancellation mechanism is required for any model in order to
cross $w=-1$.

\begin{figure}
\epsfig{figure=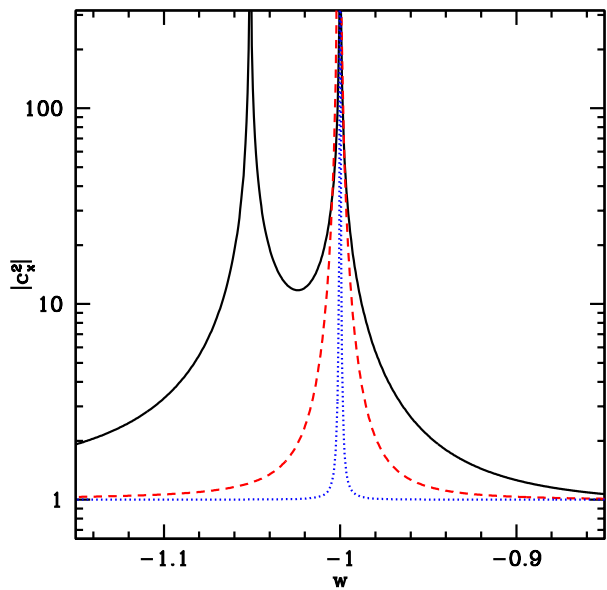,
  width=3.3in}
\caption{
We plot the
apparent sound speed $c_x^2$ defined by Eq.~(\ref{eq:cx})
for three different wave vectors, $k=1/H_0$ (black solid
line), $k=10/H_0$ (red dashed line)
and $100/H_0$ (blue dotted line). 
Although the real sound speed is just $\cs=1$,
the apparent sound speed diverges at $w=-1$ and can even
become negative.
\label{fig:quint_cs}}
\end{figure}

We also see that although the effective sound speed remains simply
$\cs=1$, this is only true if we know that there are internal relative
and non-adiabatic pressure perturbations (as well as their form). But
in general we would try to parametrise the pressure perturbation
as:
\be
\dep_\eff = c_x^2 \der_\eff +  3\HH\left(c_x^2-\ca \right) \brho_\eff 
\frac{V_\eff}{k^2} .
\label{eq:cx}
\ee
where the apparent sound speed $c_x^2$ is now a mixture of 
the real effective sound speed together with
the relative and the non-adiabatic pressure perturbations.
In this case we no longer find a simple form for the sound speed.
In Fig. \ref{fig:quint_cs} we plot $c_x^2$
as a function of the equation of state parameter $w$ for several
wave vectors $k$. As predicted, the apparent rest-frame sound speed 
diverges at the crossing. In the Quintom case the effective
perturbations do not vanish at $w=-1$.

We think that the lessons learned from the Quintom model are applicable
also to more general models with multiple fields, non-minimally coupled
scalar fields, brane-world models and other modified gravity models that
can be represented by an effective
dark energy fluid. In all these models, as in the Quintom case, $w_\eff=-1$ is not
a special point in their evolution. There is no reason to expect that the 
model will adjust its behaviour at this point, as the crossing of the phantom divide is
incidential. In representing these kinds of models with an effective fluid
we therefore expect that the perturbations will not vanish, and that the
apparent sound speed would have to diverge.

One important difference between the Quintom model and more general modified
gravity models is that the latter seem to require generically a non-zero 
anisotropic stress as $\phi\neq\psi$ \cite{ks_mg}. In this case $\cs=0$ does not
allow crossing the phantom divide, since $D_\times$ in Eq.~(\ref{eq:V_cross})
contains an additional non-zero term due to the anisotropic stress.
This reinforces our view that the Quintom-like crossing, where all 
perturbations stay non-zero and $\dep_\eff$ remains finite in spite of 
divergences in most conventionally expected terms, is more generic.
This means that there is no obvious way to predict the form of 
$\dep_\eff$ in general for modified gravity models. On the positive
side, if the dark energy is not just a cosmological constant, and
if we are able to measure $\dep_\eff$ then we may hope that this will
provide us with clues about the physical mechanism that is causing the
accelerated expansion of the universe.

\section{Conclusions}

As long as the cosmological data indicates the presence of a dark energy
with an effective equation of state $p\approx-\rho$ it will be necessary to
consider models with the same equation of state. In general, we will have
to allow the equation of state to cross the phantom divide, $w=-1$. Even
though such a fluid model may not be viable at the quantum level, it is 
possible that this behaviour is only apparent, or due to a modification
of General Relativity or the existence of more spatial dimensions. In analysing
the data we therefore have to be able to use general self-consistent models
at the level of linear perturbation theory.

In this paper we have studied the behaviour of the perturbations in general
perfect fluid models close to $w=-1$. We have shown that although models with purely
adiabatic perturbations cannot cross $w=-1$ without violating important
physical constraints (like causality or smallness of the perturbations),
it is possible to rectify the situation by allowing for non-adiabatic sources
of pressure perturbations. However, the parametrisation of $\dep$ in terms
of the rest-frame perturbations of the energy density cannot be used as this
frame becomes unphysical at $w=-1$. By parameterising $\dep$ instead in any
other frame the divergencies are avoided. 

We also computed all quantities in the Quintom model which provides an
explicit example of the above mechanism. In this model, even though the
propagation speed of sound waves remains finite and constant, the additional
internal and relative pressure perturbations lead to an apparent sound
speed which diverges. It is only the sum of all contributions to $\dep$
which remains finite.

\begin{figure}
\epsfig{figure=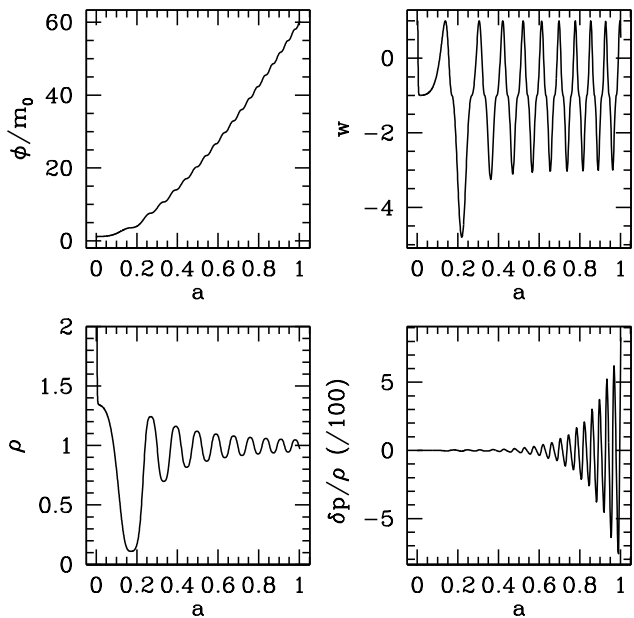,width=3.3in}
\caption{The phaxion: The sign in front of the kinetic term of a normal
scalar field is flipped every time $\dot{\phi}$ passes through zero
(so that $\dot{\phi}\geq0$). As potential we use 
$V(\phi)=1+\cos(\phi/m_0)$ which is bounded both from above and below.
The field then moves to large values (top left graph), while $w$
oscillates around $-1$ (top right). The energy density $\rho$ also
oscillates so that the field mimics an effective
cosmological constant (bottom left). As the phantom divide is crossed 
with zero slope, there is no divergence in the pressure perturbation
(bottom right), but the time-dependent effective mass can lead to
strong growth of the perturbations in spite of $\cs=1$.
\label{fig:phaxion}}
\end{figure}

\begin{figure}
\epsfig{figure=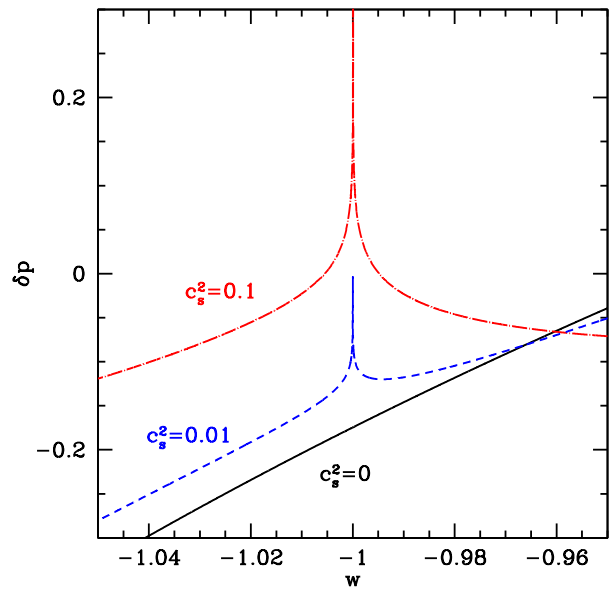,width=3.3in}
\caption{The pressure perturbation $\dep$ for three fluids with different
rest-frame sound speed, $\cs=0.1$ (top curve, red dash-dotted line),
$\cs=0.01$ (middle curve, blue dashed line) and $\cs=0$ (lowest, solid line).
Only the last case does not
have a logarithmic divergence in the pressure perturbation at the phantom
divide.
\label{fig_cs0}}
\end{figure}

A more speculative conclusion is that it seems difficult for 
``fundamental'' fields to cross $w=-1$ as their apparent rest-frame 
sound speed, defined through Eq.~(\ref{eq:csrest}), would generally be the actual
propagation speed of their perturbations, which must remain smaller than the
speed of light. From the discussion in section \ref{sec:non_ad} we learn that
fields with a well defined rest-frame and sound speed have
only two routes to phantom crossing:
\begin{itemize}
\item {\em $dw/da=0$ at crossing:} This looks rather fine-tuned as the field
needs to be aware of the presence of the barrier. On the other hand, a normal
minimally coupled scalar field reaches $w=-1$ when $\dot{\phi}=0$, and it does
so with zero slope. A scenario where phantom crossing of this kind is realised
could be built by changing the sign in front of the kinetic
term whenever $\dot{\phi}=0$. As an example, a cosine potential then leads to
a ``phaxion'' scenario, see Fig. \ref{fig:phaxion}. 
On average, the phaxion behaves like
a cosmological constant, but oscillates around $w=-1$. However, it is unclear
how to construct such a scenario in a covariant way.
\item {\em vanishing sound speed:} A field with a vanishing sound speed can
also avoid the logarithmic divergence of $\dep$. This requires either a
coupling between $1+w$ and $\cs$ or else the sound speed might be zero at
all times, see Fig.~\ref{fig_cs0}. In the latter case 
the pressure of the field would need to remain close to $-\rho$ at all times 
in order to prevent the small-scale perturbations from growing too quickly.
Also, the sound speeds needs to be {\em exactly} zero, otherwise the logarithmic
divergence reappears. This may require a symmetry enforcing $\cs=0$
as even a very small but non-zero sound speed would render this scenario
inviable.
\end{itemize}
In view of these difficulties, the Quintom family of models, although more
complicated than a single field, may prove to be the conceptually simplest
way to cross $w=-1$ with a model defined through an action. They also illustrate
that the effective pressure perturbation may provide a kind of fingerprint
of the mechanism behind the accelerated expansion of the universe if it can
be measured.

\begin{figure}
\epsfig{figure=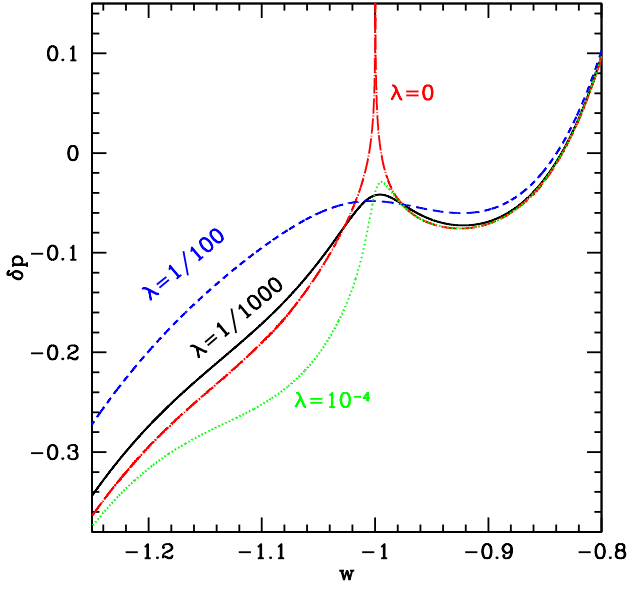,width=3.3in}
\caption{The pressure perturbation for a fluid example with $\cs=0.1$.
The red dash-dotted curve shows the standard case with the logarithmic
divergence at $w=-1$. The blue (dashed), black (solid) and green (dotted)
curve use a regularised adiabatic sound speed $\tilde{c}_a^2$ which does
not diverge. The black curve with $\lambda=1/1000$ provides a reasonable
fit. Larger values smooth too much, while smaller values start
to follow the divergence and exhibit a temporary instability in the solution
for $V$.
\label{fig_interpolation}}
\end{figure}

Concerning the use of phantom-crossing fluids for the purpose of data analysis,
it is straightforward to implement them by just avoiding the usual parametrisation
of the pressure perturbations in terms of the rest-frame sound speed.
However, there does not seem to be a canonical way to choose the pressure
perturbations if one is not allowed to use the dark energy rest-frame. The aim
for the far future will be to directly measure the pressure perturbations of
the dark energy in order to gain insight into the physical origin of the
phenomenon. For now, we have to ensure that the definition of $\dep$ does not lead
to unphysical situations, while preserving the usual parametrisation in terms
of the rest-frame sound speed as much as possible far away from the phantom
divide. Maybe the simplest way out is to regularise the adiabatic sound speed
$\ca$ which appears in Eq.~(\ref{eq:dp_rest}) because of the gauge transformation 
into the rest-frame. While any finite choice of $\dep$ is a physically acceptable 
choice, it is preferable to modify Eq.~(\ref{eq:dp_rest}) in a minimal way so
that the usual interpretation of $\cs$ is preserved for $w\neq-1$.
We propose to use
\be
\tilde{c}_a^2 = w - \frac{\dot{w} (1+w)}{3 \HH [(1+w)^2+\lambda]} 
\label{eq:intca}
\ee
where $\lambda$ is a tuneable parameter which determines how close to $w=-1$
the regularisation kicks in. A value of 
$\lambda\approx 1/1000$
should work reasonably well, as shown in Fig.~\ref{fig_interpolation}.
In this case $\dep$ is well-defined and there are no divergencies appearing
in the perturbation equations (\ref{eq:pert_cs_d}) and (\ref{eq:pert_cs_v}).
Although the differences in $\dep$ for the different choices of $\lambda$ in
Fig.~\ref{fig_interpolation} look important, we have to remember that, firstly,
the perturbations in the dark energy are normally small (especially close to the
phantom divide) and that secondly they are only communicated to the other
fluids via the gravitational potential $\psi$. As in Fig.~\ref{fig-2b},
we find also here that the dark energy perturbations are subdominant.
Computing the CMB power spectrum for $\lambda=1/100$ and $\lambda=10^{-4}$
we find a relative difference in the $C_\ell$ of less than 1\% on large scales,
which rapidly drops to $10^{-6}$ by $\ell=50$, well below cosmic variance.
We expect detectable differences only for strongly clustering dark energy
with a sound speed that is close to zero or even negative. Although it is reassuring
that directly measureable quantities are not sensitive to the precise
value of $\lambda$ in Eq.~(\ref{eq:intca}), this also shows that it
will be very difficult to distinguish between different models of dark
energy if $w\approx-1$.

\begin{acknowledgments}

M.K. and D.S. are supported by the Swiss NSF.
It is a pleasure to thank Bruce Bassett, Camille Bonvin, 
Takeshi Chiba, Anthony Lewis, Andreas Malaspinas, Norbert Straumann, 
Naoshi Sugiyama, Jochen Weller, Peter Wittwer and especially
Filippo Vernizzi, Ruth Durrer and Luca Amendola for helpful
and interesting discussions.

\end{acknowledgments}

\begin{appendix}

\section{Equivalence between scalar fields and fluid models \label{app:scalar}}

The aim of this appendix is to show that at the level of first-order
perturbation theory a scalar field behaves just like a non-adiabatic
fluid with $\cs=1$. To this end we decompose the scalar field into
a homogeneous mode $\p(t)$ and a perturbation $\delta\p(k,t)$. At the
background level we find

\bea
\brho &=& \frac{1}{2a^2}\dot{\p}^2+V\left( \p \right)\\
\bap&=& \frac{1}{2a^2}\dot{\p}^2-V\left( \p \right)
\eea
and the equation of conservation is just the equation of motion,
\be
\ddot{\p}+2 \HH \dot{\p}+a^2\frac{dV}{d\p}=0 .
\ee
The adiabatic sound speed is defined as:
\be
\ca=\frac{\dot \bap}{\dot\brho}=\frac{-\frac{3aH}{a^2}\dot{\p}^2 -2\frac{dV}{d\p}\dot{\p}}{-\frac{3aH}{a^2}\dot{\p}^2}=1+\frac{2a}{3H}\frac{\frac{dV}{d\p}}{\dot{\p}}
\ee
(and we remind the reader that $\dot{a}/a=\HH=a H$).
The perturbed energy momentum tensor is:
\bea
-\delta T_{0}^{0}&=&\delta\rho =
\frac{1}{a^2}\dot{\p}\dot{\delta\p}-\frac{1}{a^2}\dot{\p}^2\Psi+\frac{dV}{d\p}\delta\p\label{derho}\\
\delta T_{i}^{i} &=& \delta p = \frac{1}{a^2}\dot{\p}\dot{\delta\p}-\frac{1}{a^2}\dot{\p}^2\Psi-\frac{dV}{d\p}\delta\p\label{dep}\\
-ik\delta T_{0}^{i} &=& ik\delta T_{i}^{0} = \frac{k^2}{a^2}\dot{\p}\delta\p
=\brho V\label{detheta}
\eea
where we wrote $\Psi$ for the gravitational potential $\psi$ in order to avoid
confusions with the scalar field variables (only in this appendix).

In order to derive the rest frame sound speed $\cs$ of the scalar field
we use equation (\ref{eq:dp_rest})
\be
\delta p = \cs\delta\rho +
\frac{3aH}{k^2}\left(\cs-\ca\right)\brho V
\ee
and express everything in terms of scalar field quantities. We find
\bea
&&\frac{1}{a^2}\dot{\p}\dot{\delta\p}-\frac{1}{a^2}\dot{\p}^2\Psi-\frac{dV}{d\p}\delta\p=\nonumber\\
&&=\cs\left(\frac{1}{a^2}\dot{\p}\dot{\delta\p}-\frac{1}{a^2}\dot{\p}^2\Psi+\frac{dV}{d\p}\delta\p
+\frac{3aH}{a^2}\dot{\p}\delta \p\right)-\nonumber\\
&&-\frac{3aH}{a^2}\left(1+\frac{2a}{3H}\frac{1}{\dot{\p}}\frac{dV}{d\p}\right) \dot{\p}\delta\p
\eea
which after some algebraic manipulations turns into
\bea
&&\frac{1}{a^2}\dot{\p}\dot{\delta\p}-\frac{1}{a^2}\dot{\p}^2\Psi+\frac{dV}{d\p}\delta\p
+\frac{3aH}{a^2}\dot{\p}\delta \p=\nonumber\\
&&=\cs\left(\frac{1}{a^2}\dot{\p}\dot{\delta\p}-\frac{1}{a^2}\dot{\p}^2\Psi+\frac{dV}{d\p}\delta\p
+\frac{3aH}{a^2}\dot{\p}\delta \p\right) .
\eea
Therefore $\cs=1$.

Now let us derive the equation of motion for the scalar field perturbations
from the perturbation equation for a perfect fluid, Eq.~(\ref{d_pert}),
\be
\dot{\delta}=-\left(1+w\right)\left(\theta-3\dot\Psi\right)
-3aH\left(\frac{\delta p}{\brho} -w\delta \right)
\ee
which can be rewritten as
\be
\dot{\delta\rho}+3aH\left(\brho+\bap\right)\delta =
3\left(\brho+\bap\right)\dot{\Psi}-\rho V \label{cont}
\ee

Expressing the time derivative $\delta\rho$ in terms of scalar field quantities,
\bea
\dot{\delta\rho}&=&-\frac{2aH}{a^2}\dot{\p}\dot{\delta\p}+\frac{1}{a^2}\ddot{\delta\p}\dot{\p}-\frac{2aH}{a^2}\dot{\p}\dot{\delta\p}-\frac{dV}{d\p}\dot{\delta\p}- \label{deltarhodot} \\
&&-\frac{1}{a^2}\dot{\p}^2\dot{\Psi}+\frac{6aH}{a^2}\dot{\p}^2\Psi+2\dot{\p}\frac{dV}{d\p}\Psi+\frac{dV}{d\p}\dot{\delta\p}+\frac{d^2V}{d\p^2}\dot{\p}\delta\p\nonumber
\eea
and doing likewise with the other terms,
\bea
3aH\left(\delta\rho +\delta p\right) &=& \frac{6aH}{a^2}\dot{\p}\dot{\delta\p}-\frac{6aH}{a^2}\dot{\p}^2\Psi\label{drho+dp} \\
3\left(\brho+\bap\right)\dot\Psi &=& \frac{3}{a^3}\dot{\p}^2\dot\Psi\label{rho+p} \\
\brho V &=&\frac{k^2}{a^2}\dot\p \delta\p\label{V}
\eea
we can insert all these expressions into Eq. (\ref{cont}) and obtain finally
\be
\ddot{\delta\p}+2aH\dot{\delta\p}+a^2\left( \frac{d^2V}{d\p^2}
+\frac{k^2}{a^2}\right)\delta\p=4\dot\p\dot\Psi -2a^2 \Psi\frac{dV}{d\p}\label{cons}
\ee
which is indeed the equation of motion for $\delta\p$ in the conformal
Newtonian gauge (see e.g.~\cite{hu_lecture}).

\section{Effective perturbations in two barotropic fluids \label{app:2baro}}

We consider first two barotropic perfect fluids with constant
equation of state parameters,
\bea
\bar p_{1}=w_{1} \bar\rho_{1}, \quad w_{1}>-1; \\
\bar p_{2}=w_{2} \bar\rho_{2}, \quad w_{2}<-1 .
\eea
We can define the effective quantities as those appearing in the
sum of the two energy-momentum tensors,

\bea
\brho_\eff &=& \brho_{1} + \brho_{2} \label{effrho}\\
\bap_\eff &=& \bap_{1} + \bap_{2}\\
w_\eff &=& \frac{w_{1}\brho_{1} + w_{2}\rho_{2}}{\brho_{1}+\brho_{2}}\\
\delta_\eff &=& \frac{\brho_{1}\delta_{1} + \brho_{2}\delta_{2}}{\brho_{1}+\brho_{2}} \\
\theta_\eff&=&\frac{\left( 1+w_{1} \right)\brho_{1}\theta_{1}+\left( 1+w_{2}
  \right)\brho_{2}\theta_{2}}{\left( 1+w_{1} \right)\brho_{1}+\left( 1+w_{2}
  \right)\brho_{2}} . \label{tet_effective}
\eea
The system above is characterized by four variables $\delta_1$, $\delta_2$,
$\theta_1$, $\theta_2$. In order to have a complete mapping, we need to
introduce two more variables which express two internal degrees of
freedom \cite{MaWa}:
\bea
S_{12}&=& \frac{\delta_1}{1+w_1} - \frac{\delta_2}{1+w_2}  \label{rel_del}\\
\Delta\theta_{12} &=& \theta_{1} - \theta_{2} \label{t_1_2}
\eea
called respectively, relative entropy perturbation and the relative velocity
of the two fluids.

We can now calculate the perturbation equation for the effective fluid from
\be
T^{\mu}_{_{(eff)} \nu;\mu}=T^{\mu}_{_{(1)} \nu;\mu}+T^{\mu}_{_{(2)} \nu;\mu}=0
\ee

The $T^{\mu}_{0;\mu}$ and $T^{\mu}_{\nu;\mu}$ components give the perturbation equations:
\bea
&&\brho_{1} \dot\delta_{1} + \brho_{2} \dot\delta_{2} - 3 \dot\phi \left[ \left( 1+w_{1} \right) \brho_{1}+ \left( 1+w_{2} \right) \brho_{2} \right] + \nonumber\\
&&+ \left[ \left( 1+w_{1} \right) \brho_{1} \theta_{1} + \left( 1+w_{2} \right) \brho_{2} \theta_{2} \right] + \nonumber \\
&&+3\HH \left[ \delta p_{1} -w_{1} \brho_{1} \delta_{1}+ \delta p_{2} -
  w_{2} \brho_{2} \delta_{2} \right] =0
\label{d1+d2}
\eea
the derivative of $\dot\delta_\eff$ is:
\bea
\dot\delta_\eff &=& \frac{\brho_{1}\dot\delta_{1}+\brho_{2} \dot\delta_{2}}{\brho_{1} + \brho_{2}} - 3\HH\frac{w_{1}\brho_{1}\delta_{1}+ w_{2}\brho_{2}\delta_{2}}{\brho_{1}+\brho_{2}}+\nonumber\\
&&+3\HH\frac{w_{1}\brho_{1}+w_{2}\brho_{2}}{\brho_{1}+\brho_{2}}\frac{\brho_{1}\delta_{1}+\brho_{2}\delta_{2}}{\brho_{1}+\brho_{2}}\label{delta_dot}
\eea
inserting the last one into Eq.~(\ref{d1+d2}) and remembering Eqs.~(\ref{effrho}) to (\ref{rel_del}), we have:
\bea
\dot\delta_\eff &=& -\left( 1+w_\eff \right) \left( \theta_{\eff} -3\dot\psi \right) \nonumber \\
&&-3\HH \left( \frac{\delta p_\eff}{\brho_\eff} - w_\eff\delta_\eff \right) \label{d_eff_2}
\eea
where $\delta p_\eff = \delta p_{1} + \delta p_{2}$. This is just the equation
(\ref{d_pert}) for the effective quantities.

For the second perturbation equation we have:
\bea
&&\left( 1+w_{1} \right) \brho_{1} \dot\theta_{1} +\left( 1+w_{2} \right) \brho_{2} \dot\theta_{2} +\nonumber\\
&&+4\HH \left[ \left( 1+w_{1} \right) \brho_{1} \theta_{1} + \left( 1+w_{2} \right) \brho_{2} \theta_{2} \right]+ \nonumber \\
&&-3\HH  \left[\left( 1+w_{1} \right)^{2} \brho_{1} \theta_{1} + \left( 1+w_{2} \right)^{2} \brho_{2} \theta_{2} \right]+\nonumber\\
&&-k^{2}\psi \left[ \left( 1+w_{1} \right) \brho_{1} +\left( 1+w_{1} \right) \brho_{1} \right] +\nonumber \\
&&- k^{2} \left( \delta p_{1} + \delta p_{2} \right) =0 \label{t1+t2}
\eea
the derivative of $\theta_\eff$ is:
\bea
\dot\theta_\eff&=&\frac{\left( 1+w_{1} \right) \brho_{1} \dot\theta_{1} + \left( 1+w_{2} \right) \brho_{2} \dot\theta_{2}}{\left( 1+w_{1} \right) \brho_{1} +\left( 1+w_{1} \right) \brho_{1}} +\nonumber \\
&&-3\HH \frac{ \left( 1+w_{1} \right)^{2} \brho_{1} \theta_{1} + \left( 1+w_{2} \right)^{2} \brho_{2} \theta_{2}}{\left( 1+w_{1} \right) \brho_{1} +\left( 1+w_{1} \right) \brho_{1}}+ \nonumber \\
&&+3\HH\frac{\left( 1+w_{1} \right) \brho_{1} \theta_{1} + \left( 1+w_{2} \right) \brho_{2} \theta_{2} }{\left( 1+w_{1} \right) \brho_{1} +\left( 1+w_{1} \right) \brho_{1}}\times \nonumber \\
&&\times\frac{\left( 1+w_{1} \right)^{2} \brho_{1} + \left( 1+w_{2} \right)^{2} \brho_{2}}{\left( 1+w_{1} \right) \brho_{1} +\left( 1+w_{1} \right) \brho_{1}}
\eea
again inserting the last one into Eq.~(\ref{t1+t2}) and remembering
Eqs.~(\ref{effrho}) to(\ref{tet_effective}) and (\ref{rel_del}), we obtain the expression for the second perturbation equation:
\bea
\dot\theta_\eff&=&-\HH\left( 1-3w_\eff \right) -\frac{\dot w_\eff}{1+w_\eff} \theta_\eff \nonumber \\
&&+k^{2} \left[ \frac{\delta p_\eff/\brho_\eff}{1+w_\eff} +\psi \right] = 0 \label{t_eff_2}
\eea
Again, this is the same equation as the one for a single perfect fluid, Eq.~(\ref{t_pert}).
The difference to the general case is that
now the pressure perturbations are fixed by the barotropic nature of the
two fluids.
Starting from the generic expression
\be
\dep_\eff = \dep_{1} + \dep_{2}=w_{1}\brho_{1}\delta_{1}+w_{2}\brho_{2}\delta_{2}
\ee
we find that the effective pressure perturbation is
\be
\dep_\eff = \ca \delta_\eff - \frac{1}{3\HH} \frac{ \left(w_{2}-w_{1} \right) \dot\brho_{1} \dot\brho_{2}}{\dot\brho_\eff } S_{12} \label{p_rel_app}
\ee
where $\ca$ is the adiabatic sound speed and is given by
\bea
\ca &=&
\frac{\dot\bap_\eff}{\dot\brho_\eff}=\frac{w_{1}\dot\brho_{1}+w_{2}\dot\brho_{2}}{\dot\brho_{1}+\dot\brho_{2}}=\nonumber \\
&=& \frac{w_{1} \left( 1+w_{1} \right)\brho_{1} +w_{2} \left(
  1+w_{2} \right) \brho_{2}}{\left( 1+w_{1} \right)\brho_{1}+ \left( 1+w_{2}
  \right)\brho_{2}} .
\eea

The second term appearing in Eq.~(\ref{p_rel_app}) can be considered as
the relative pressure perturbation due to the relative
motion:
\be
\delta p_{\rel} = - \frac{\left( w_{2} - w_{1} \right) \dot\brho_{1} \dot\brho_{2}}{3\HH \dot{\brho}_\eff} S_{12}
\ee

A single barotropic fluid has the pressure perturbation $\dep=\ca \der$. For the
two-fluid case we find an additional part coming from the relative perturbations
of the two fluids.
%
%
%
%
The new variable $S_{12}$  given by Eq.~(\ref{rel_del}) is a gauge invariant relative entropy perturbation \cite{MaWa}.

The time evolution of $S_{12}$ is given simply by:
\be
\dot S_{12}= -\Delta\theta_{12} ,
\label{S_12}
\ee
and contains the fourth variable $\Delta \theta_{12}$, which is a relative velocity
perturbation and evolves according to
\bea
\dd_t \Delta\theta_{12}&=&-\HH\Delta\theta_{12}+3\HH\left( w_{1}+w_{2}-c^{2}_{a} \right) \Delta\theta_{12} + \nonumber \\
&&+k^{2} \left(  w_{1}+w_{2}-c^{2}_{a} \right) S_{12}+ \nonumber \\ 
&&+3\HH \left(  w_{1}-w_{2} \right) \theta_\eff + k^{2}\frac{ w_{1}-w_{2}}{1+w_\eff} \label{Dt_eff}
\eea

\section{Effective perturbations in the Quintom model \label{app:2fluids}}

Perturbations in barotropic fluids with constant $w<0$ grow very rapidly due
to the imaginary sound speed, $\ca=w<0$. A realistic model crossing the phantom
divide needs therefore to be composed of fluids with non-adiabatic fluctuations
and a positive sound speed. In the case of the Quintom model, we are dealing
with two fluids with $\cs=1$. As in appendix \ref{app:2baro}, we define the
effective quantities via eq. (\ref{effrho})~-~eq. (\ref{t_1_2}): using these equations we find the relations between the variables of the two fluids and the variables of the
effective fluid:
\bea
\delta_{1} &=& \frac{1+w_{1}}{1+w_\eff}\delta_\eff-\nonumber\\
&&-\frac{\left(1+w_{1}\right) \left( 1+w_{2} \right) 
  \left(w_\eff-w_{1}\right)}{\left( 1+w_\eff\right) \left(
  w_{2}-w_{1}\right)}S_{12} \label{d_1}\\
\delta_{2} &=& \frac{1+w_{2}}{1+w_\eff}\delta_\eff-\nonumber\\
&&-\frac{\left(1+w_{1}\right) \left( 1+w_{2} \right)
  \left(w_{2}-w_\eff\right)}{\left( 1+w_\eff\right) \left(
  w_{2}-w_{1}\right)}S_{12} \label{d_2}\\
\theta_{1}&=&\theta_\eff+\frac{\left( 1+w_{2}\right) \left( w_\eff-w_{1}\right)}{\left( 1+w_\eff\right) \left(
  w_{2}-w_{1}\right)}\Delta\theta_{12}\label{t_1}\\
\theta_{2}&=&\theta_\eff+\frac{\left( 1+w_{1}\right) \left( w_\eff-w_{2}\right)}{\left( 1+w_\eff\right) \left(
  w_{2}-w_{1}\right)}\Delta\theta_{12} \label{t_2}
\eea

What we need to evaluate again in the Quintom model is the effective 
pressure perturbation, because all the other terms are the same. Taking the
sum of the pressure perturbation defined in the rest-frame, we have:
\bea
\dep_\eff &=& \csa\der_{1}+\csb\der_{2}+ \nonumber\\
&&+3\HH\left(1+w_{1} \right) \left(
    \csa-w_{1}\right)\brho_{1}\frac{\theta_{1}}{k^2}+\nonumber \\
&&+3\HH\left(1+w_{2} \right) \left( \csb-w_{2}\right)\brho_{2}\frac{\theta_{2}}{k^2}\label{dp_eff_rest}
\eea
inserting the Eqs.~(\ref{d_1}) - (\ref{t_2}) in Eq. (\ref{dp_eff_rest}) we find:
\bea
&&\dep_\eff= \cseff\der_\eff+\dep_{\rel} +\dep_{\nad}
\nonumber\\
&&+ 3\HH\left(\cseff-\ca \right) \left(
1+w_\eff\right)\brho_\eff\frac{\theta_\eff}{k^2}
\eea
where $\cseff$ is the effective rest-frame sound speed;
$\dep_{\rel}$ is the relative pressure perturbation and $\dep_{\nad}$ is the non
adiabatic contribution to the pressure perturbation; they are:
\bea
\cseff&=&\frac{\csa\left( 1+w_{1} \right) \brho_{1}+\csb\left( 1+w_{2} \right) \brho_{2}}{\left( 1+w_{1} \right) \brho_{1}+\left( 1+w_{2} \right) \brho_{2}}\\
\dep_{\rel}&=&\frac{\left( w_{2} - w_{1} \right) \dot\brho_{1}
  \dot\brho_{2}}{3\HH \dot{\brho}_\eff} S_{12}\\
\dep_{\nad}&=&-\left[\left(\csa-\csb\right) + \left(
  w_{2}-w_{1}\right)
  \right]\frac{\dot\brho_{1}\dot\brho_{2}}{3\HH\dot\brho_\eff}\times\nonumber\\
&&\times\left[S_{12}+ \frac{3\HH}{k^2}\Delta\theta_{12}\right]
\eea

The total number of degrees of freedom remains the same when we change from
the ``two fluid'' to the ``single effective fluid'' picture. In both cases
we have four variables. What changes is the way these variables interact.
In the two fluid case, the interaction proceeds through the gravitational
potential $\psi$. In the single effective fluid picture, the additional
degrees of freedom become internal and appear through additional contributions
to the pressure perturbation $\dep$.

\end{appendix}


\begin{thebibliography}{}
\bibitem{sn1} A. G. Riess et al., Astronomical J. {\bf 116}, 1009 (1998).
\bibitem{sn2} S. Perlmutter et al., Astrophys. J. {\bf 517}, 565 (1999).
\bibitem{cora} P.S. Corasaniti et al., Phys. Rev. D {\bf 70} 083006 (2004).
\bibitem{wlim} D.N. Spergel et al., astro-ph/0603449 (2006).
\bibitem{phantom} U. Alam et al., Mon. Not. Roy. Astron. Soc. {\bf 354} 275 (2004).
\bibitem{caldwell} R.R. Caldwell, Phys. Lett. B {\bf 545}, 23 (2002).
\bibitem{crossing} R.R. Caldwell and M. Doran, Phys. Rev. D {\bf 72}, 043527 (2005).
\bibitem{cht} S.M. Carroll, M. Hoffmann and M. Trodden, Phys. Rev. D {\bf 68}, 023509 (2003).
\bibitem{Cline:2003gs}
J.~M.~Cline, S.~Jeon and G.~D.~Moore,
Phys.\ Rev.\ D {\bf 70}, 043543 (2004).
\bibitem{para} L. Perivolaropoulos, JCAP {\bf 0510}, 001 (2005).
\bibitem{nojiri} S. Nojiri and S.D. Odintsov, Gen. Rel. Grav. {\bf 38}, 1285 (2006).
\bibitem{vli} M. Li, B. Feng and X. Zhang, JCAP {\bf 0512}, 002 (2005).
\bibitem{vik2} A. Anisimov, E. Babichev and A. Vikman, JCAP {\bf 0506}, 006 (2005).
\bibitem{kosh} I.Ya. Aref'eva, A.S. Koshelev and S.Yu. Vernov, Phys. Rev. D {\bf 72}, 064017 (2005).
\bibitem{hu_lecture} W. Hu, lecture notes, astro-ph/0402060 (2004).
\bibitem{mabe} C.P. Ma and E. Bertschinger, Astrophys. J. {\bf 455}, 7 (1995).
\bibitem{vik1} A. Vikman,  Phys. Rev. D {\bf 71}, 023515 (2005).
\bibitem{ah} A. Adams et al., hep-th/0602178 (2006).
\bibitem{bcd} C. Bonvin, C. Caprini, R. Durrer, astro-ph/0606584 (2006).
\bibitem{bedo} R. Bean and O. Dor\'e, Phys. Rev. D {\bf 69}, 083503 (2004).
\bibitem{KS} H. Kodama and M. Sasaki, Progr. Theor. Phys. Suppl. {\bf 78}, 1 (1984).
\bibitem{quintom1} B. Feng, X. Wang and X. Zhang, Phys. Lett. B {\bf 607}, 35 (2005).
\bibitem{quintom2} W. Hu, Phys. Rev. D {\bf 71}, 047301 (2005).
\bibitem{qpert} G.-B. Zhao et al., Phys. Rev. D {\bf 72}, 123515 (2005).
\bibitem{MaWa} K. A. Malik and D. Wands, JCAP {\bf 0502}, 007 (2005).
\bibitem{ks_mg} M. Kunz and D. Sapone, astro-ph/0612452 (2006).
\bibitem{Carroll} Sean M. Carroll, An Introduction to General Relativity SpaceTime and Geometry (Addison Wesley, 2004)
\bibitem{Ruth} R. Durrer. Gauge Invariant Cosmological perturbation Theory,
  Fund. Cosmic Phys. {\bf 15}, 209 (1994).
\end{thebibliography}
\end{document}